\documentstyle[twoside,fleqn,espcrc2,epsfig]{article}
\def\beq{\begin{equation}}
\def\eeq{\end{equation}}
\def\bea{\begin{array}}
\def\eea{\end{array}}
\def\beqa{\begin{eqnarray}}
\def\eeqa{\end{eqnarray}}

\def\u1{{U(1)}}
\def\su2{{SU(2)}}
\newcommand{\re}{\relax{\rm I\kern-.18em R}}

\newcommand{\AmS}{{\protect\the\textfont2
  A\kern-.1667em\lower.5ex\hbox{M}\kern-.125emS}}

\title{Gauge Theory and String Theory;\\
An Introduction
to the AdS/CFT Correspondence\thanks{Plenary talk at
{\it LATTICE 99} held in Pisa, Itay from June 29 to July 3, 1999. 
To appear in the Proceedings of the
conference.}
}
\author{Hirosi Ooguri\thanks{On leave of absence
from University of California, Berkeley.}
\\Caltech -- USC Center for Theoretical Physics, 
Mail Stop 452-48 \\
California Institute of Technology, 
Pasadena, CA 91125, USA}
\begin{document}
\begin{abstract}

\end{abstract}
\maketitle

\section{Introduction}

In this talk, I would like to show you some of the
recent developments in superstring theory, in particular
the relation between gauge theory and string theory.
String theory was originally invented as a theory of
hadrons, but it was superseded by the gauge theory.
It then found its employment in quantum gravity.
Now it seems that string theory and gauge theory
are meeting again, and I hope this new direction
will provide an interesting arena where lattice
gauge theorists and string theorists can interact and
exchange ideas, benefiting both.

Let me begin my talk by posing a question:

\medskip

\centerline{{\it What is the string theory?}}

\medskip

\noindent
Until recently, one of the serious defects of the string theory
had been that we did not know what it is. We did not have a
definition of the theory. This was an unpleasant situation
since there was a logical possibility that the string theory
might not exist after all, even as a purely theoretical framework,
let alone the possibility of describing the real world.
Fortunately we now know that string theory exists in certain
situations, and it is what I would like to tell you today.
But before we get into that, let me tell you what we knew before.

The original ``definition'' of the string theory was entirely
in terms of the perturbative expansion. We knew that the theory,
if exists, should contain oscillating strings, which in the limit
of the vanishing coupling constant
$g_{string} \rightarrow 0$, freely propagate in spacetime. 
We also knew how to compute string amplitudes using the 
perturbative expansion in $g_{string}$. In superstring, each
term in the perturbative expansion was known to be finite.
However this, by itself, cannot be a complete definition. 
The perturbative series is not convergent in most of the
cases. So, even though superstring theory was proposed as
the unified theory including gravity, we could not use 
it to study effects in strong gravitational
fields to address mysteries of quantum gravity. 
There was a concern that there may be some unknown strong
coupling phenomenon which makes the theory ill-defined.

\medskip

\centerline{{\it Does string theory exist?}}

\medskip

Recently a remarkable correspondence between
gauge theory and string theory was discovered
\cite{malda,gkp,witten} (see \cite{review} for
a review)
and it has partially resolved this problem.
It also realized the earlier expectation that 
the 't Hooft large-$N$ limit of gauge theory
is a string theory \cite{thooft}. The realization, however,
came with a twist --- that the string theory
and the gauge theory live in different dimensions. 
The correspondence was discovered in the study of
black hole in string theory. 

\section{Black $p$-branes in string theory}

It has been recognized for a long time that 
quantizing gravity is not a straightforward task.
There are many aspects of quantum gravity 
which we do not know how to deal with 
in the standard field theory method. 
If string theory truly unifies quantum mechanics 
and general relativity, string theory should be able 
to address these puzzles of quantum gravity. 

One of the important questions in quantum gravity 
is the information paradox of black hole. 
The problem is roughly as follows. 
The black hole is characterized 
by the presence of an event horizon, which separates 
its interior region from an outside observer.
Now suppose we throw some object 
into the black hole.
Once the object passes the horizon,  
we won't be able to access to the information
carried by the object. 
In classical general relativity, the information is not lost, 
but is merely hidden behind the horizon. 
In quantum mechanics, however, the situation is different.
It was showed by Hawking \cite{hawking}, based on the semi-classical
approximation, that the black hole emits pure thermal 
radiation, 
namely with the radiation with maximum randomness. 
So it seems that the information carried by
the object 
is completely dissipated as the thermal radiation. 
The loss of information may not be a problem in 
classical physics, but it is a big problem in quantum 
mechanics since it means that we cannot maintain 
the quantum coherence under the time evolution.
It violates the basic axioms of quantum mechanics.  
This is the information paradox of quantum black hole. 

String theory in fact has a large class 
of black hole solutions.
They were called black $p$-branes, where 
$p$ is an integer and it can be $0, 1, 2$ etc. 
The $0$-brane is the standard black hole, 
which is a point source.
The spacetime becomes flat when one goes away 
from the center, but it is strongly curved 
near the center and there is a horizon. 
When $p = 1$, we have the black $1$-brane. 
It is a string-like source. 
It is spread in one-spatial dimensions. 
Combined with the time direction, the source 
looks like a two-dimensional surface. 
The $2$-brane is a membrane. 
In general, the $p$-brane is extended in 
$p$-spatial directions and $1$-time direction.

So the string theory indeed have black hole solutions. 
To address the question of the information paradox, 
one needs to know what happens 
if we throw  a string into the black $p$-brane. 

There are two ways to approach this problems. 
One is to study the propagation of the string
in the black $p$-brane geometry. 
There is, however, another way,
that is to use the collective coordinates of
the $p$-brane. 
In this approach, we first identify
intrinsic degrees of freedom of the $p$-brane, 
which describe the motion and the fluctuation of the brane. 
One can then try to formulate the problem 
as interactions between the string and 
the collective coordinates of the $p$-brane. 
This approach was taken by Polchinski \cite{polchinski}, 
and he called this description of the $p$-brane as 
D-brane. In a certain situation, 
in particular when we study physics at low energy, 
this D-brane description tells you 
that the dynamics of the $p$-brane is described 
by a gauge theory in $(p+1)$ dimensions

So we have these two descriptions of the same object. 
In one way of viewing at it, we have strings moving
in the background geometry of black $p$-brane. 
In another way of viewing at it, we have 
the collective coordinates of the $p$-brane, 
which is the $(p+1)$-dimensional gauge theory, 
interacting with strings.
The equivalence of these two descriptions is 
the origin of the connection between 
gauge theory and string theory.
In the past few years, various evidences 
have emerged in support of the idea that 
gauge theory and quantum gravity are closely related.
At the same time, 
the gauge theory description has provided us 
very strong computational tools to study string theory.

\section{AdS/CFT correspondence}

The evidences in support of the correspondence between
gauge theory and string theory
 have crystallized in the work of Maldacena. 
He formulated the conjecture that
superstring theory in a curved $10$-dimensional space,
which is $5$-dimensional anti-de Sitter space ($AdS_5$)
times $5$-dimensional sphere ($S^5$),
is equivalent to
a gauge theory in $4$ dimensions with $N=4$ supersymmetry. 
The anti-de Sitter space is a homogeneous space with  
negative curvature, which I will describe in more detail below. 
The gauge theory with $N=4$ supersymmetry is
a conformal field theory (CFT).
Thus the conjecture by Maldacena is called
the AdS/CFT correspondence.

To formulate the conjecture,
Maldacena looked at the black $3$-brane.
The D-brane description suggests that 
the low energy physics of the $3$-brane is described 
by the gauge theory, in this case the 4-dimensional
gauge theory with $N=4$ supersymmetry.
This description becomes exact 
in the low energy limit. 
He then noticed that, in precisely the same limit, 
the black hole description also becomes nice;
in this limit,  the region near 
the event horizon of the 3-brane is amplified.
The geometry near the horizon is that of 
$AdS_5 \times S^5$. In this way, the essence of
the correspondence between 
gauge theory and string theory 
has been extracted. 

\begin{figure}[htb]
\begin{center}
\epsfxsize=1.8in\leavevmode\epsfbox{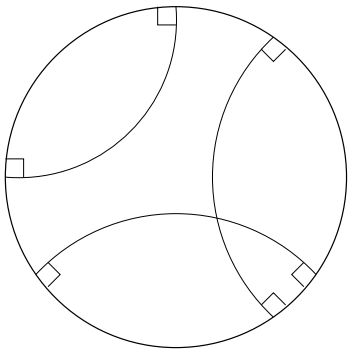}
\end{center}
\caption{Poincare disc model of the hyperbolic space.
Geodesics are represented by semi-circles which intersects
with the boundary in the right angle.}
\label{fig1}
\end{figure}

\begin{figure}[htb]
\begin{center}
\epsfxsize=2in\leavevmode\epsfbox{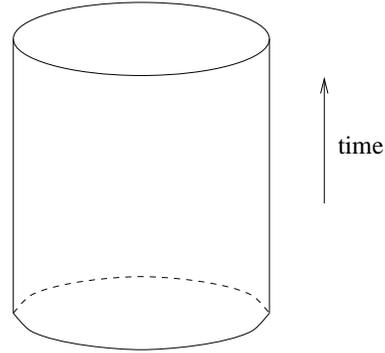}
\end{center}
\caption{$AdS_5$ is obtained by adding one time direction 
to the 4-dimensional
hyperbolic space.}
\label{fig2}
\end{figure}

Let me describe what the anti-de Sitter space is. 
The 5-dimensional anti-de Sitter space ($AdS_5$) 
has $4$ spatial dimensions and $1$ time. 
The spacelike section of $AdS_5$ is simply 
the hyperbolic space of Lobachevsky and Bolyai. 
It is historically the first counter-example 
to Euclid's 5th postulate about parallel lines. 
In Figure 1, I show Poincare's disk model 
of the hyperbolic space.
In this model, geodesics are represented by 
semi-circles which are intersecting with the boundary 
in the right angle. In our case, the disk is $4$ dimensional, 
so its boundary is 3-dimensional sphere. 
$AdS_5$
is obtained by simply adding $1$ time direction to this.
You may view it like a solid cylinder as in Figure 2.  
The boundary of $AdS_5$ is 4 dimensional space  of 
$R \times S^3$,  
and it is identified as a spacetime for the gauge theory. 

More precisely, the conjecture states that Type IIB superstring
theory (consisting only of closed strings with the same
chirality in both left and right moving sectors on the worldsheet)
is equivalent to the ${\cal N}=4$ supersymmetric gauge
theory in 4 dimensions with gauge group $SU(N)$. 
Type IIB superstring on $AdS_5 \times S^5$ has three dimensionful
parameters, $l_{AdS}, l_{string}$ and $l_{Planck}$. 
By the stringy generalization of Einstein's equation,
the radius of $S^5$ is required to be the same as
the curvature radius $l_{AdS}$ of $AdS_5$.
The string length $l_{string}$ characterizes
the size of the zero point oscillation of the string,
$i.e.$ (string tension)$=l_{string}^{-2}$. Finally the Planck
length $l_{Planck}$ in 10 dimensions is expressed in the
combination of the string coupling constant $g_{string}$
and the string length as $l_{Planck} 
= g_{string}^{1/4} l_{string}$.
The theory is then characterized by two dimensionless combinations
of these, for example $l_{AdS}/l_{string}$ and $L_{AdS}/l_{Planck}$. 
On the other
hand, the gauge theory also has two parameters, the
size of the gauge group $N$ and the gauge coupling constant
$g_{gauge}$. (Since the ${\cal N} = 4$ theory is ultraviolet finite,
$g_{gauge}$ is really a constant.) These two sets of parameters
are related as
$$  {l_{AdS} \over l_{string}} = (g_{gauge}^2 N)^{1/4},~~~
{l_{AdS} \over l_{Planck}} = N^{1/4}. $$
In particular, if we take the limit $N \rightarrow \infty$
keeping $g_{gauge}^2 N$ finite, quantum gravity effects in 
$AdS_5 \times S^5$ are suppressed and 
we have strings freely propagating in the spacetime.
This realizes the idea that the 't Hooft large-$N$ limit
of gauge theory is a tree-level string theory. 

After Maldacena's work,  the conjecture was formulated 
in a more precise manner and 
various tests of this conjecture have been made.  
The AdS/CFT correspondence states that 
the two quantum theories, namely
the string theory in $10$ dimensions and 
the gauge theory in $4$ dimensions are equivalent.
This means that, first of all, 
the Hilbert space of the two theories must be identical. 
The Hilbert space of the string theory contains
gravitons and strings propagating in the AdS space,
and black holes and various black p-branes. 
On the other hand, the Hilbert space of the gauge theory
is constructed from the gauge invariant observables
such as the field strength of the gauge field. 
So there has to be a dictionary between the two. 
In fact, we have succeeded in identifying 
what the gravitons correspond to in the gauge theory side, 
and we have also understood some aspects of 
black holes from the gauge theory point of view.  
The dictionary of the two Hilbert spaces
is still incomplete, and  there are many things 
to be understood here.

The quantum theories are characterized by 
the operator algebras on the Hilbert spaces. 
Thus the equivalence of the two theories implies 
that the correlation functions of the two theories must be the same. 
After the conjecture was formulated, 
various computations have been done on both side,
and many non-trivial agreements have been found. 
String theory computations also give 
various theoretical prediction about the gauge theory, 
and these are currently being examined using 
the gauge theory method. 

There are two surprising features of the AdS/CFT correspondence. 
The first surprise is that string and the gauge theory live 
in different spacetime dimensions. 
I should point out that this is not the same 
as the Kaluza-Klein reduction. 
In the Kaluza-Klein mechanism, one curls up 
a part of the spacetime into a tiny circle. 
If we ignore field fluctuation on the circle 
(or compact manifolds, in general), 
the dimensionality of the space is reduced. 
Here we are not truncating a theory. 
The equivalence states that the string theory is 
fully equivalent to the gauge theory, without any reduction. 
The second surprise is that string theory contains 
gravity and the gauge theory doesn't. 
These two surprises
seem to be related to an earlier observation 
on quantum gravity by 't Hooft \cite{thoofthol} and 
Susskind \cite{susskind}. 
They suggested, based on properties of black hole, 
that in quantum gravity the information 
of the theory can be stored in lower dimensions. 
This idea is called Holography of Quantum Gravity. 
It seems that the AdS/CFT correspondence is 
an explicit realization of this idea.

\section{Summary and Outlook}

Let me summarize what have been accomplished. 
I think that the most important fact coming out from 
this development is that we now have a complete definition 
of the string theory in certain cases. 
For superstring on $AdS_5 \times S^5$,
the 4-dimensional gauge theory with $N=4$ supersymmetry
can in principle give the non-perturbative definition. 
In addition, the AdS/CFT correspondence 
realizes the idea that the 't Hooft large-$N$ limit
of the $SU(N)$ gauge theory is string theory.
It also realizes the idea that quantum gravity is holographic. 

I would like to close this talk by pointing out 
some future directions. 
Since we know string theory and gauge theory 
are equivalent, we may try to use string theory 
to do difficult computations in gauge theory or vice versa, 
and thereby doubling our theoretical knowledge. 
In fact string theory computations have shown  
us various new results about 
gauge theories with conformal symmetry. 
For applications to QCD, to
go beyond the qualitative analysis, 
we need to understand string dynamics 
in the curved background better \cite{wittenqcd,gross}, 
and I hope there will be some progress in this direction. 

One should also hope to learn much more about 
quantum gravity using gauge theory. 
Some aspects of quantum black holes have been studied 
using the gauge theory method, 
in particular the microscopic derivation of 
the black hole entropy by Strominger and Vafa
\cite{vafa}. 

The correspondence has so far been limited to 
the case when string theory is on a space 
which is asymptotically anti de Sitter. 
It is desirable to extend this to other geometry 
such as asymptotically flat space.
More generally, one should hope to find 
a formulation of string theory independently
of its background geometry.

The main aim of string theory research is still 
a search for the unified theory. 
In the course of this research, we have 
found that string theory is also useful 
to study various gauge theories. I hope that
useful collaborations between lattice
gauge theorists and string theorists would emerge
from this correspondence.

\section*{Acknowledgments}
I would like to thank the organizers
of {\sl LATTICE 99} for the very stimulating conference
and for their hospitality.  
This research  was supported in part by 
NSF grant PHY-95-14797, DOE grant DE-AC03-76SF00098,
and the Caltech Discovery Fund.

\end{document}